\documentclass[reprint,amssymb,amsmath,aip,cha]{revtex4-1}
\usepackage[dvips]{graphicx}
\usepackage[dvips]{epsfig,color}
\usepackage{dcolumn}
\usepackage{bm}% bold math

\begin{document}
\title{The role of orbital order in the stabilization of the ($\pi,0$) 
ordered magnetic state in a minimal two-band model for iron pnictides}
\author{Sayandip Ghosh}
\email{sayandip@iitk.ac.in}
\author{Avinash Singh}
\affiliation{Department of Physics, Indian Institute of Technology Kanpur -
208016}
\begin{abstract}
Spin wave excitations and stability of the ($\pi,0$) ordered magnetic state 
are investigated in a minimal two-band itinerant-electron model for iron
pnictides. Presence of hopping anisotropy generates a strong ferro-orbital order
in the $d_{xz}$ and $d_{yz}$ Fe orbitals. The orbital order sign is as observed
in experiments. The induced ferro-orbital order strongly enhances the
spin wave energy scale and stabilizes the magnetic state by optimizing the
strength of the emergent AF and F spin couplings through optimal band fillings
in the two orbitals. The calculated spin-wave dispersion is in quantitative
agreement with neutron scattering measurements. Finite inter-orbital Hund's
coupling is shown to further enhance the spin wave energies state by coupling
the two magnetic sub-systems. A more realistic two-band model with less hopping
anisotropy is also considered which yields not only the circular hole pockets,
also correct ferro-orbital order and emergent F spin coupling.
\end{abstract}
\maketitle

\section{Introduction}
The iron pnictides exhibit a typical phase diagram \cite{Zhao,Nandi} in which
the parent compound goes through a tetragonal-to-orthorhombic structural phase
transition (at $T_S$) and spin ordering transition (at $T_N$). Upon doping, both
transitions are suppressed and superconductivity emerges. Single crystal neutron
scattering experiments show that Fe moments align antiferromagnetically (AF)
along the $a$ direction and ferromagnetically (F) along the $b$
direction \cite{Goldman}, so that the magnetically ordered state can be viewed
as a ($\pi,0$) ordered spin density wave (SDW) state. Inelastic neutron
scattering (INS) experiments \cite{Zhao2,Diallo,Ewings} yield well-defined
spin-wave excitations up to the F zone boundary  $\bm q$=($0,\pi$) on an energy
scale $\sim$ 200 meV.
 
Several weak coupling models with Fermi surface (FS) nesting have been proposed
\cite{Eremin,Brydon,Knolle} 
to account for the observed magnetic order. Although
these models can explain low-energy magnetic excitations, they fail at higher
energies and suggest that spin waves enter the particle-hole continuum at high
energies and become over-damped Stoner-type excitations \cite{Knolle}. The
recently observed existence \cite{Ewings} of
spin-wave excitations even above $T_N$ is also contrary to this weak coupling
nesting picture, within which there is no difference between moment melting and
moment disordering temperatures. In fact, LDA
calculations \cite{Nakamura,Skornyakov} suggest that Fe onsite
interaction $U$ is
comparable to Fe $3d$ bandwidth (W) indicating that iron pnitides are
moderately correlated materials. 

Apart from the magnetic excitations, angle-resolved photoemission spectroscopy
(ARPES) \cite{Shimojima,Yi,Jensen} and X-ray linear dichroism (XLD)
experiments \cite{Kim} have clearly revealed the existence of orbital order in
these
materials. In the magnetic state, the Fe $d_{yz}$
band is shifted up relative to the $d_{xz}$ band \cite{Yi,Kim}, causing
electron density difference between the two orbitals. This type of orbital
ordering was previously proposed \cite{Lv,Lee,Chen,Lv2} to explain
experimentally observed in-plane anisotropic
behavior like anisotropy in magnetic exchange coupling \cite{Zhao2}, transport
properties \cite{Tanatar,Chu}, FS structure \cite{Yi}, and electronic structure
\cite{Chuang}.

Is there a significant correlation between this observed orbital ordering and
stability of the ($\pi,0$) magnetic state? In this paper, we will investigate
the effect of orbital ordering on the SDW state stability within a minimal
two-band itinerant-electron model. We will study how the variations in orbital
disparity affects the induced AF and F spin couplings. This provides a
microscopic understanding of the role of orbital ordering on the stability of
the ($\pi,0$) ordered SDW state in the relevant intermediate coupling regime.

The minimal two-band model proposed earlier
by Raghu \textit{et al.} \cite{raghu} gave Fermi surface structure consistent
with LDA calculations at half-filling. Although nesting between
hole and electron Fermi pockets yields low critical value of $U$ for
($\pi,0$) ordering, it has three major shortcomings: (i) no F spin
coupling is generated due to nesting as shown by the vanishing spin wave energy
at the F zone boundary \cite{jpcm2,knolle2}, whereas
INS experiments yield maximum spin wave energy, (ii) no orbital ordering is
obtained in this model for electron filling corresponding to nesting condition,
and (iii) Fermi surface nesting is relevant only in weak coupling limit,
whereas
pnictides are in intermediate coupling regime ($U\sim W$). At intermediate
coupling, nesting is not very relevant for
magnetic ordering as evidenced by the observation of a stable ($\pi,0$) state
in one-band Hubbard model \cite{jpcm1}.

Evidently, modifications are required to this two-band model and therefore
investigation of models and
mechanisms beyond nesting become relevant in order to
reproduce experimentally observed orbital ordering as well as the spin wave
dispersion. The present study is an investigation in this direction. As hopping
anisotropy has been shown to yield orbital order in the
($\pi,0$) state \cite{Phillips}, we will
first consider the extreme hopping anisotropy case within a minimal two-band
model (Section 2) and investigate the consequences on spin couplings and spin
wave dispersion (Section 3), to bring out in
a physically transparent manner how
the resulting orbital order enhances the F and AF spin couplings in
stabilizing the ($\pi,0$) structure and yielding the observed features of the
spin wave
dispersion. Guided by the above investigations, Section 4 describes
modifications to the two-band model to obtain circular hole pockets and
orbital order in agreement with experiments, and also emergence of required F
spin coupling. This comprehensive approach of simultaneously keeping account of
the Fermi surface as well as emergent spin couplings and spin wave dispersion
 provides important physical insight into further extension to a three-band
model.

\section{Minimal two-band model}
\begin{figure}
\begin{center}
 \vspace*{-0mm}
\hspace*{0mm}
\psfig{figure=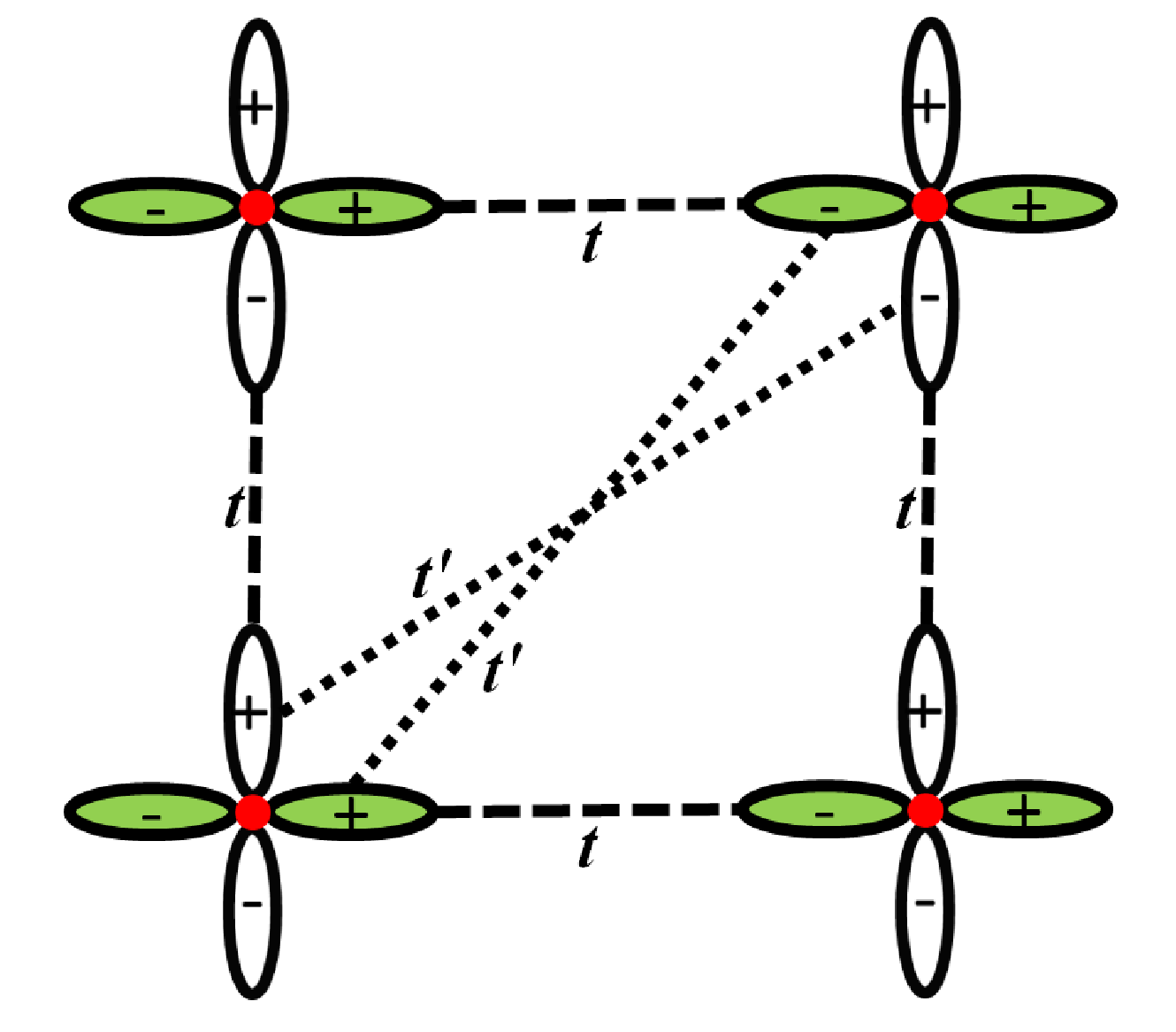,width=55mm,angle=0}
\vspace*{-0mm} 
\end{center}
\caption{\label{model}
Effective hopping parameters $t_{ij}^{\mu\nu}$ in the minimal two-band model
involving $d_{xz}$ (filled) and $d_{yz}$ (empty) orbitals, referred to as
$\alpha$ and $\beta$
respectively.}
\end{figure}

The iron pnictides have a quasi two-dimensional structure with layers of FeAs
stacked along the \textit{c} axis. Among the five Fe $3d$ orbitals, only
$d_{xz}$ and $d_{yz}$ contribute to orbital ordering (due to $C_4$ symmetry
of others) and we retain only these two in our model. The
hybridization of Fe $3d$ orbitals with themselves as well as through the As $3p$
orbitals lying above and below the square plaquettes formed by the Fe atoms
leads to hopping parameters in our two-orbital model. The hopping amplitudes are
shown in Fig. \ref{model} in which the $d_{xz}$ and $d_{yz}$ orbitals are taken
to be extended along $x(a)$ and $y(b)$ direction respectively since the cores of
Fe $d_{xz}$ and $d_{yz}$ Wannier Functions extend towards the direction of
magnetic ordering \cite{Lee}. Although hybridization between the orbitals can
lead to finite $t_\pi$ (i.e. $\pi$ overlap), for simplicity we consider
$t_\sigma$ (i.e. $\sigma$ overlap) only in our model. However, orbital order
in our model persists as long as there is hopping anisotropy i.e. $t_\sigma$ is
larger than $t_\pi$. A finite intra-orbital next-nearest-neighbor (NNN) hopping
$t^{\prime}$, expected due to presence of Fe-As-Fe path along plaquette
diagonals, is also
included. Finite $t^{\prime}$ plays a very important role in stabilizing the SDW
ordering. 
 
We start with the two-orbital Hubbard model Hamiltonian:
\begin{equation}
H = -\sum_{\langle ij\rangle \mu\nu\sigma} t_{ij}^{\mu\nu} 
(a_{i\mu\sigma}^\dagger a_{j\nu\sigma} + a_{j\nu\sigma}^\dagger
a_{i\mu\sigma}) + \sum_{i \mu } U_{\mu } n_{i \mu \uparrow} n_{i \mu
\downarrow}
\end{equation}
where $i$, $j$ refer to lattice sites; $\mu$, $\nu$ are the orbital
indices, $t_{ij}^{\mu\nu}$ are the hopping terms as shown in Fig. \ref{model},
and $U_\mu$ are the intra-orbital Coulomb interactions. The role of
inter-orbital density interaction and Hund's coupling will be discussed later.

In this model, the two orbitals are decoupled, with non-magnetic state 
dispersions $\varepsilon_{\alpha}({\bf k})$=$-2t\cos k_x-4t^{\prime}\cos
k_x\cos k_y $ for the $d_{xz}$ and $\varepsilon_{\beta}({\bf k})$=$-2t\cos
k_y-4t^{\prime}\cos k_x\cos k_y $ for the $d_{yz}$ orbitals. Although the
dispersions are different, the energy bands are degenerate in the non-magnetic
state wherein
\textit{x} and \textit{y} directions are equivalent. This degeneracy is
important to
satisfy point-group symmetry conditions. It is noteworthy that the $d_{xz}$
band along $\Gamma-Y$ and $d_{yz}$ along $\Gamma-X$ are degenerate in energy,
as indeed observed in ARPES studies \cite{Yi} on $\rm BaFe_2 As_2$ in the
non-magnetic
state above the ordering temperature.

In the ($\pi,0$) ordered SDW state, the Hartree-Fock (HF) level Hamiltonian
matrix is expressed in a composite two-orbital ($\alpha$  $\beta$),
two-sublattice (A B) basis as:
\begin{equation}
H_{\rm HF}^{\sigma} ({\bf k}) = \left [ \begin{array}{cccc} -\sigma
\Delta_\alpha  \;\; & \varepsilon_{\bf k}^{x} + \varepsilon_{\bf k}^{xy} & 0 & 0
\\ 
\varepsilon_{\bf k}^{x} + \varepsilon_{\bf k}^{xy} \;\; & \sigma \Delta_\alpha &
0 & 0 \\ 
0 \;\; & 0 & - \sigma \Delta_\beta + \varepsilon_{\bf k}^{y} & \varepsilon_{\bf
k}^{xy} \\ 
0\;\; & 0 & \varepsilon_{\bf k}^{xy} & \sigma \Delta_\beta + \varepsilon_{\bf
k}^{y} 
\end{array} \right ]
\label{Hamiltonian}
\end{equation} 
where  $\varepsilon_{\bf k}^{x(y)}$=$-2 t \cos k_{x(y)}$, $\varepsilon_{\bf
k}^{xy}$=$-4t^{\prime}\cos k_x\cos k_y$, $\sigma$=$\pm$ for the two spins. The
self-consistent exchange fields are given
by $2\Delta_\mu$=$U_{\mu} m_{\mu}$ in terms of the sublattice magnetizations
$m_\mu$ for the two orbitals $\mu=\alpha,\beta$. As the intermediate-coupling
regime will be considered throughout, 
the term ``SDW state'' is used here without any implicit weak-coupling
connotation. 

The density terms arising in the HF approximation, $U_{\mu}n_{\mu}/2 =
\Delta_{\mu} + U_{\mu}n_{\mu}^{\downarrow}$ 
for the two orbitals, have not been shown explicitly in (1). In the following,
we will take identical exchange fields
$\Delta_{\alpha}=\Delta_{\beta} \equiv \Delta$, for which we find the relative
band shift 
$U_{\alpha}n_{\alpha}^{\downarrow} - U_{\beta}n_{\beta}^{\downarrow}$ to be
quite small, and this will be absorbed in an energy offset to be introduced
later.

In iron pnictides, each iron ion has six electrons distributed among five
$3d$ orbitals. Two $e_g$ orbitals $d_{x^2-y^2}$ and $d_{3z^2-r^2}$ are
completely occupied by four electrons due to large crystal-field splitting
between $e_g$ and $t_{2g}$ states \cite{kruger}, and the three $t_{2g}$ orbitals
are partially filled by the remaining two electrons. ARPES
experiments \cite{Yi} show that the $d_{xy}$ orbital has a finite
contribution to the Fermi Surface. Therefore, the expected electron filling in
our two band model will be less than half filling, and will correspond to a
``hole doped'' condition.

\begin{figure}
\begin{center}
\vspace*{-0mm}
\hspace*{-0mm}
\psfig{figure=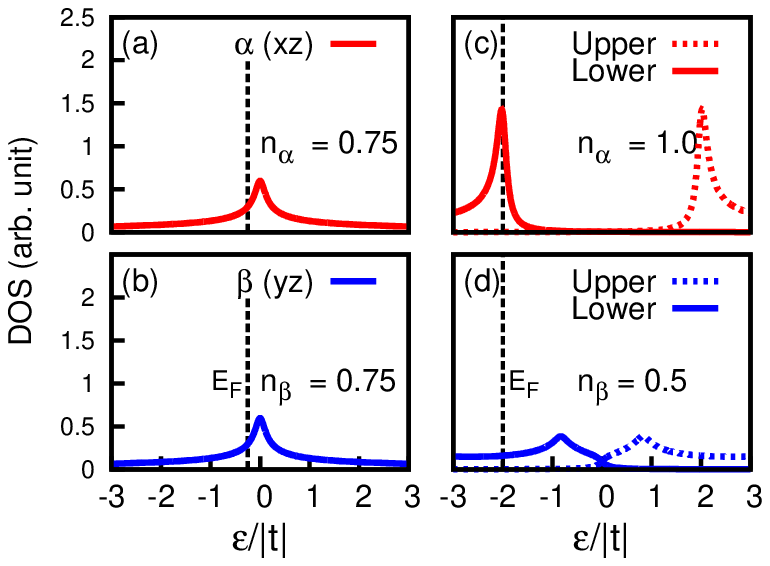,width=50mm,angle=-90}
\vspace{-0mm}
\end{center} 
\caption{\label{result2}
Calculated partial densities of state (DOS) for $\alpha$ and $\beta$
orbitals in the non-magnetic [panels (a),(b)] and magnetic [panels (c),(d)]
states. Here, $t'/t$=0.5. The
degeneracy between $\alpha$ and $\beta$ orbitals is
lifted by magnetic ordering. For a typical $E_F$ as shown,
$n_{\alpha}>n_{\beta}$. Enhanced $n_\alpha$ and reduced $n_\beta$ due to hopping
anisotropy 
result in stronger AF and F spin couplings.}
\end{figure}

The partial densities of states for the two orbitals are shown in Fig.
\ref{result2} for the non-magnetic ($\Delta$=0) and magnetic ($\Delta/|t|$=2)
cases.
While the two orbitals are degenerate in the non-magnetic state, 
the degeneracy is lifted in the magnetic state since $x$ and $y$ directions are
no longer equivalent. The hopping anisotropy together with magnetic ordering
anisotropy naturally
results in self orbital order in the magnetic state. 
For hole doping ($E_F<0$), the lower $\beta$ band becomes partially unoccupied 
whereas the lower $\alpha$ band remains nearly half-filled. 
For total electron occupation $n\approx$ 0.75 per orbital (i.e. $25\%$ hole
doping), we have $n_\alpha\approx$ 1.0 and $n_\beta\approx$ 0.5. 
This sign of ferro-orbital order ($n_\alpha - n_\beta >0$) is in agreement with 
experiments \cite{Shimojima,Yi,Jensen,Kim}. The higher value of $\alpha$ DOS
than $\beta$ DOS at Fermi energy naturally leads to more
conductivity along AF direction than F direction which agrees with experiments
\cite{Chu}.

\section{Transverse spin fluctuations}
Spin-wave excitations in this spontaneously-broken-symmetry SDW state
are obtained from the transverse spin fluctuation propagator:
\begin{equation}
[\chi^{-+} _{\rm RPA} ({\bf q},\omega)] = \frac{[\chi^0 ({\bf q},\omega)]}{{\bf
1} - [U][\chi^0 ({\bf q},\omega)]}
\label{propagator}
\end{equation}
at the RPA level. The interaction matrix $[U]$ includes $U_{\alpha}$ and
$U_{\beta}$ as diagonal matrix elements. The bare particle-hole propagator
$[\chi^{0} ({\bf q},\omega)]$ is evaluated in the composite orbital-sublattice
basis \cite{jpcm2} by integrating out the fermions in the ($\pi,0$) ordered SDW
state as:
\begin{equation}\begin{split}
[\chi^0 ({\bf q},\omega)]_{ab} = i \int \frac{d \omega'}{2 \pi} \sum_{\bf k'}
[G_{\rm HF}^{\uparrow}({\bf k'},\omega')]_{ab}\\ 
[G_{\rm HF}^{\downarrow}({\bf
k'-q},\omega'-\omega)]_{ba}
\end{split}
\end{equation}
where $[G_{\rm HF}^{\sigma}({\bf k},\omega)]$=$
[\omega {\bf 1} - H_{\rm HF}^{\sigma}({\bf k})]^{-1}$ are the HF level Green's
function in the SDW state and $a$, $b$ being the orbital-sublattice basis
indices. The spin-wave energies are obtained from the poles of
Eq. (\ref{propagator}).

\begin{figure}
\begin{center}
\vspace*{-0mm}
\hspace*{0mm}
\psfig{figure=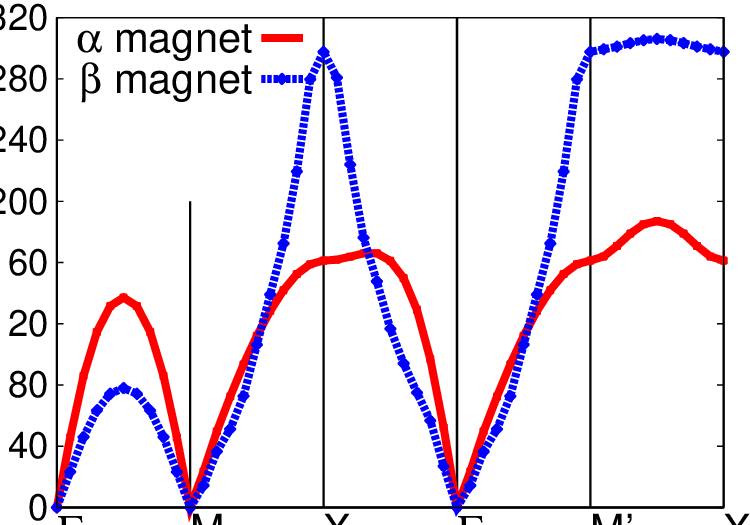,width=50mm,angle=0}
\end{center}
\caption{\label{result1}
Spin-wave dispersions for the two magnet modes $\alpha$ and $\beta$ 
along symmetry directions of the BZ for $25\%$ hole doping.
Here $t$=$-$200 meV, $t'/t$=0.5, $\Delta/|t|$=2.0, and the Fermi energy $E_{\rm
F}/|t|\approx -$2.0.}
\vspace*{-0mm}
\end{figure}

In the absence of any Hund's coupling, the two orbitals are decoupled and the
magnetic system reduces to two independent magnetic sub-systems $\alpha$ and
$\beta$ involving AF ordering in \textit{x} direction ($\alpha$ magnet) and F
ordering in \textit{y} direction ($\beta$ magnet). The spin-wave dispersions
for the two magnet modes $\alpha$ and $\beta$ are shown in Fig. \ref{result1}
along ($0,0$)$\longrightarrow$($\pi,0$)$\longrightarrow$($\pi,
\pi$)$\longrightarrow$($0,\pi$)$\longrightarrow$($\pi,\pi$)[$\Gamma
\longrightarrow M \longrightarrow X \longrightarrow M'\longrightarrow X$]. 
For $t^{\prime}$=0, the system further reduces
to independent AF and F chains, which yield zero spin wave energy for wave
vector in the respectively perpendicular directions (due
to absence of any spin coupling along those directions), implying
instabilities of the magnetic state. Fig. \ref{result1} thus highlights the
important role of finite $t^{\prime}$ and the corresponding finite inter-chain
spin couplings in stabilizing the two $\alpha$ and $\beta$ magnets with respect
to spin twisting in any direction, as indicated by positive spin wave energies
over the entire BZ.

\begin{figure}
\begin{center}
\vspace*{0mm}
\hspace*{-0mm}
\psfig{figure=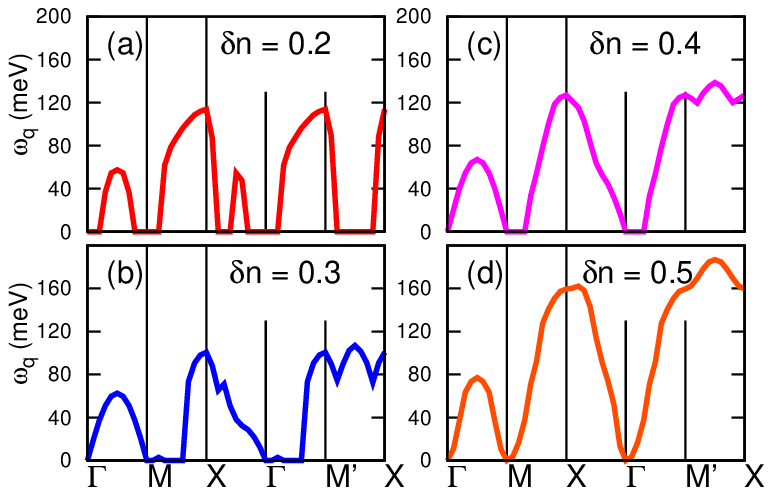,width=60mm,angle=-90}
\vspace{-5mm}
\end{center}
\caption{\label{result3}
Spin-wave energies with different orbital polarization $\delta n =
n_{\alpha}-n_{\beta}$ for $25\%$ hole doping. Crossover from negative to
positive energy modes shows strong stabilization of the AF-F state with
increasing orbital order. The hopping parameters and exchange fields are
same as in Fig. \ref{result1}.}
\end{figure}

To investigate the effect of orbital order on the SDW state stability, we
plot the spin-wave dispersions for different orbital
polarization $\delta n = n_{\alpha}-n_{\beta}$ in Fig. \ref{result3}. 
As magnetic state instability is indicated by the spin-wave energy crossing zero
and going negative, we will focus only on the lowest (out of the two $\alpha$
and $\beta$ mode)
energies. Here, we have maintained a constant SDW order parameter $\Delta$ and
total
electron filling n = 0.75 per orbital, 
and the occupations $n_{\alpha}$ and $n_{\beta}$ of the two orbitals are
controlled by introducing an energy offset $\Delta_{\alpha \beta}$ between the
two orbitals.
The orbital occupations are (a) $n_\alpha$ = 0.85, $n_\beta$ =
0.65 ($\Delta_{\alpha \beta}$=0.38), (b) $n_\alpha$ = 0.9, $n_\beta$ = 0.6
($\Delta_{\alpha \beta}$=0.29), (c) $n_\alpha$ = 0.95, $n_\beta$ = 0.55
($\Delta_{\alpha \beta}$=0.20) and (d) $n_\alpha$ = 1.0, $n_\beta$ = 0.5
($\Delta_{\alpha \beta}$=0). Figure \ref{result3} shows a strong stabilization
of the SDW state with orbital order, as seen by crossover from negative to
positive energy modes. 

The origin of this \textit{orbital-order-induced stabilization} is as follows. 
Effectively, the AF and F spin couplings are optimized by the electron density
redistribution associated with orbital order. The increased electron density in
$\alpha$ band favors AF coupling (super exchange) in the $x$ direction (NN) and
in the diagonal direction (NNN), whereas the increased hole density in $\beta$
band favors carrier-mediated F coupling in the $y$ direction (NN) and AF
coupling in the diagonal direction (NNN) \cite{Ghosh}. Thus, all the spin
couplings work together and the $\alpha$ and $\beta$ magnets both reinforce
($\pi,0$) ordering without any frustration. It is important to note here that
the magnetic and orbital orderings effectively \textit{stabilize each
other} and constitute a composite spin-orbital ordered state with $ m$$\neq$0,
$\delta n$$\neq $0. Our model thus provides a microscopic understanding of the
close relation between in-plane anisotropy, orbital order, and the SDW magnetic
order. If the inter-orbital Coulomb interaction term
$Vn_{i\alpha}n_{i\beta}$ is included in our model, it will only enhance the
orbital offset $\Delta_{\alpha \beta}$ due to orbital disparity, and therefore
enhance
the effect discussed above.

Ferro orbital order was reported in a recent study of magnetic excitations 
in iron pnictides within a degenerate double-exchange model including
antiferromagnetic superexchange interactions \cite{Phillips}. However, the sign
of the ferro orbital order reported in this work ($n_{yz}>n_{xz}$) does not
agree with experiments. Furthermore, for a realistic NN hopping value of 200
meV, their calculated spin wave energy scale of around 30 meV is well below 
the nearly 200 meV energy scale measured in INS experiments.
Ferro orbital was also reported in other multi-orbital models due to electron
correlation \cite{Quan}, anisotropic inter-orbital hopping \cite{Bascones} and
electron-lattice coupling \cite{Liu}. However, spin wave excitations and role of
orbital ordering in stabilization of ($\pi,0$) state with respect to transverse
spin fluctuations was not investigated. Moreover, crystal field splitting due to
orthorhombic distortion is necessary to stabilize orbital ordering in these
models unlike our model. 

Now, we investigate the effect of finite inter-orbital Hund's coupling $J$, 
and consider the Hamiltonian:
\begin{equation}
H = -\sum_{\langle ij\rangle \mu\nu\sigma} t_{ij}^{\mu\nu} 
(a_{i\mu\sigma}^\dagger a_{j\nu\sigma} + a_{j\nu\sigma}^\dagger
a_{i\mu\sigma}) 
- \sum_{i \mu \nu} U_{\mu \nu} {\bf S}_{i \mu} \cdot {\bf S}_{i
\nu}
\end{equation}
where the interaction matrix elements $U_{\mu \nu}$=$U_{\mu}$  for
$\mu$=$\nu$ and $U_{\mu \nu}$=$2J$ for $\mu$$\ne$$\nu$. 
The self-consistent exchange fields are now given by
$2\Delta_\alpha$=$U_{\alpha} m_{\alpha} + J m_{\beta}$ and 
$2\Delta_\beta$=$U_{\beta} m_{\beta} + J m_{\alpha}$. 

\begin{figure}
\begin{center}
\vspace*{0mm}
\hspace*{-0mm}
\psfig{figure=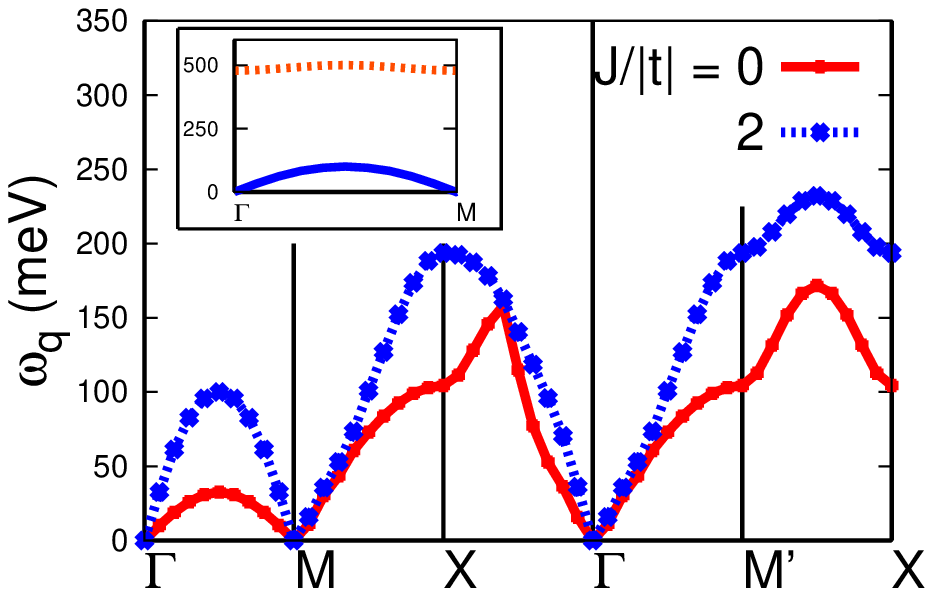,width=45mm,angle=-90}
\end{center}
\caption{\label{result5}
Strong enhancement in the spin-wave energies on including the Hund's coupling. 
Here, $t$=$-$200 meV, $t'/t$=0.3, $\Delta/|t|$=2.0, and hole doping 25\%.}
\vspace{-0mm} 
\end{figure}

The transverse spin fluctuation propagator now includes both $U$ and $J$ ladders
at RPA level. Figure \ref{result5} shows the spin-wave dispersion with and
without Hund's coupling, displaying a strong enhancement of spin-wave energies
with Hund's coupling, as expected since now the $\alpha$ and $\beta$ magnets are
coupled.

For $J$=$0$, there are two independent Goldstone modes, 
as the $\alpha$ and $\beta$ magnets are independent. When finite Hund's coupling
is included, the two modes get coupled, leading to a single Goldstone mode
corresponding to
``in-phase'' fluctuations (acoustic mode). The ``out-of-phase'' mode is now
converted
to an optical branch which rapidly becomes significantly gapped even at small
values of $J$. The spin-waves for the acoustic and optical modes (solid and
dotted respectively) are shown in the inset [Fig. \ref{result5}]. 

The calculated spin-wave energy scale ($\sim$200 meV) agrees well with neutron
scattering experiments for $|t|$=200 meV and $U$ in the intermediate coupling
range ($U\sim W$). The value of U ($\sim$1-2 eV) agrees with LDA
calculations \cite{Yang}. Furthermore, the spin-wave dispersion shows a peak at
the F zone boundary which is consistent with experiments. For $\Delta/|t|=$2.0,
the SDW state effective gap $\sim$ 400 meV is well above the maximum spin wave
energy. Thus, contrary to other itinerant models \cite{Knolle}, spin wave
excitations in our model do not rapidly dissolve into the particle-hole
continuum, as indeed not observed experimentally up to energies of 200 meV. 

\begin{figure}
\begin{center}
\vspace*{-2mm}
\hspace*{-0mm}
\psfig{figure=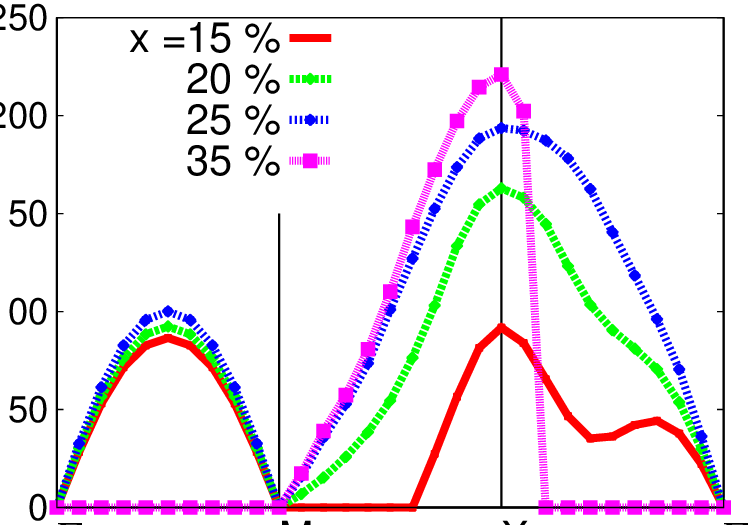,width=50mm}
\end{center}
\caption{\label{fig6}
Variation of spin wave dispersion with hole doping, showing strong suppression
of AF (F) spin couplings at high (low) doping. 
Here $t$=$-$200 meV, $t'/t$=0.3, and $\Delta$=$J$=2$|t|$.}
\vspace{-0mm} 
\end{figure}

The spin wave energy scales in the $\Gamma$-M (AF) and M-X (F) directions 
are indicative of the AF and F spin coupling strengths.
The hole doping dependence of spin wave dispersion (Fig. \ref{fig6}) effectively
shows the evolution of emergent spin couplings.
At high hole doping, the AF spin coupling is strongly suppressed due to electron
density depletion in the AF band, whereas the F spin coupling is optimized due
to enhanced hole doping in the F band. On the other hand, at low hole doping,
the AF spin coupling gets saturated as the AF (lower) band is filled, whereas
the F spin coupling is strongly weakened. 

Maximum SDW state stability is seen for about $25\%$ hole doping.
If this corresponds to the parent compound, then electron
doping (reduction in hole doping from this level) results in crossover
to negative-energy modes, indicating destabilization of the SDW state. 
This is in agreement with the observed rapid decrease of magnetic
ordering temperature in iron pnictides with electron doping \cite{Zhao}.

\begin{figure}
\begin{center}
\vspace*{-2mm}
\hspace*{-0mm}
\psfig{figure=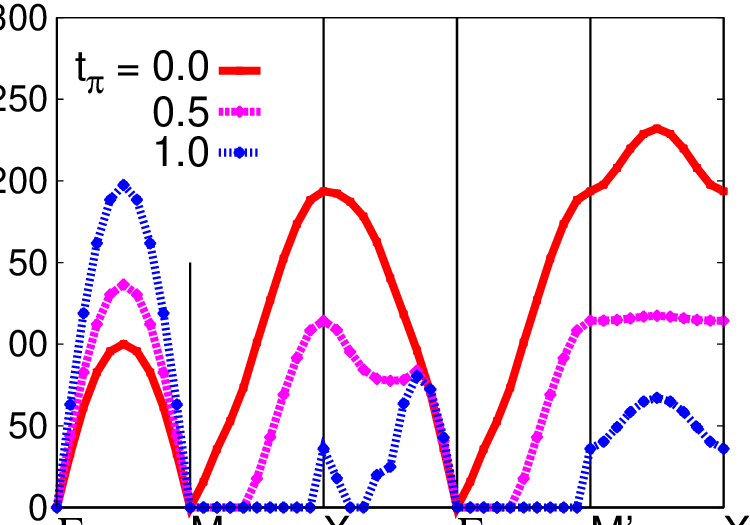,width=50mm}
\end{center}
\caption{\label{fig7}
Strong enhancement in spin wave energy scale and SDW state stability with
increasing hopping anisotropy. 
Here $t$=$-$200 meV, $t'/t$=0.3, $\Delta$=$J$=2$|t|$, and hole doping 25\%.}
\vspace{-0mm} 
\end{figure}

Spin waves were studied earlier \cite{jpcm2} in a two-band model with isotropic
hopping and
therefore no orbital order. For somewhat larger interaction strength
($\Delta/|t|)$=3) and hole doping
($\sim$40\%), the spin wave energies are qualitatively similar as in Fig.
\ref{result5}.  
However, for lower interaction strength and hole doping, increasing hopping
anisotropy significantly enhances the spin wave energy scale and SDW state
stability, as
shown in Fig. \ref{fig7}. This confirms that the emergent AF and F spin
couplings become stronger due to
the electron density redistribution and orbital order in the two orbitals,
corresponding to enhanced electron and hole densities in the $d_{xz}$ and
$d_{yz}$ Fe orbitals, respectively.

\begin{figure}
\begin{center}
\vspace*{0mm}
\hspace*{-0mm}
\psfig{figure=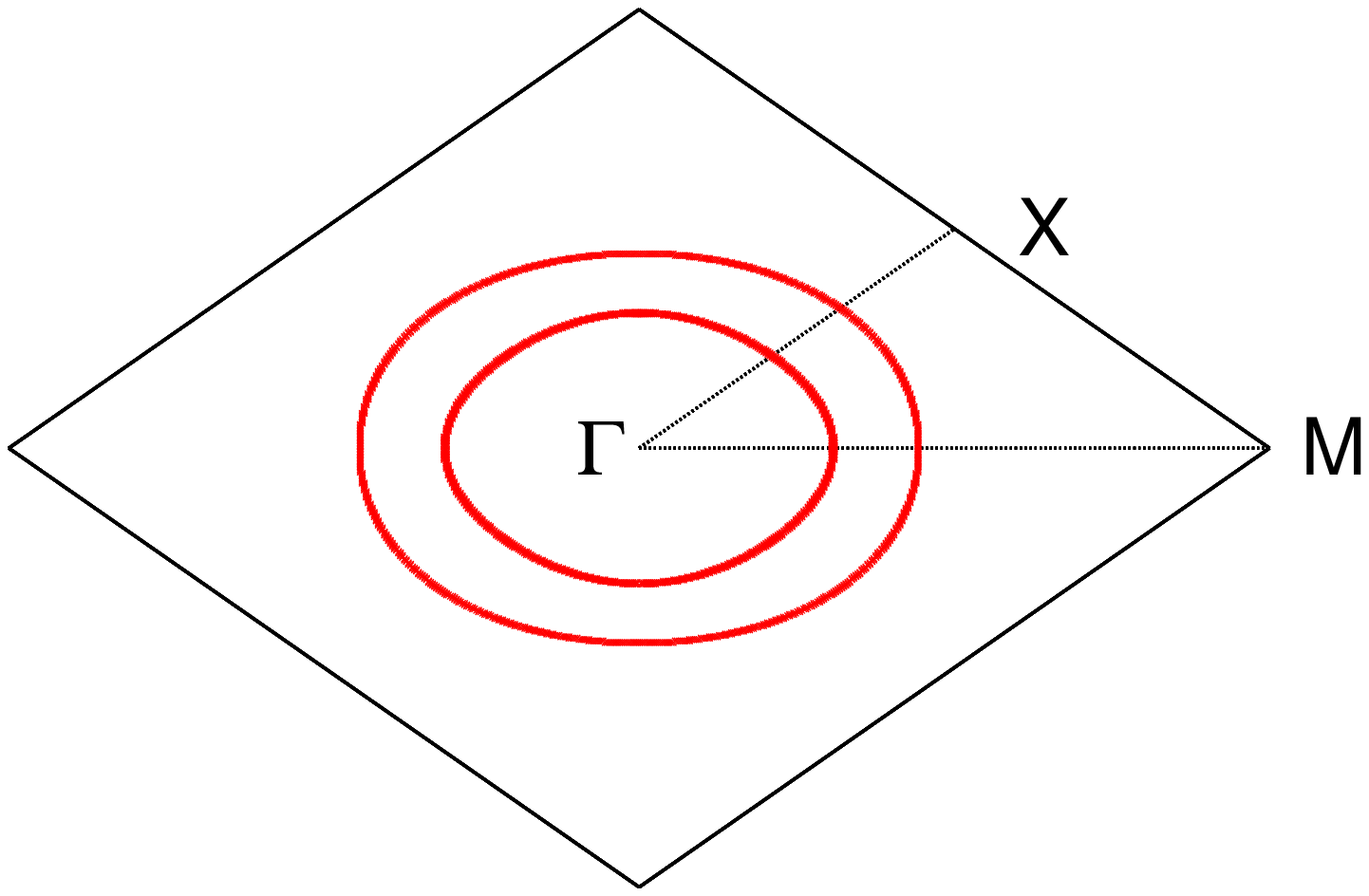,height=50mm,width=50mm,angle=270}
\end{center}
\caption{\label{fig8}
Fermi surface in the folded BZ for $t_1$=$-1.0$, $t_2$=$0.5$,
$t_3$=$t_4$=$-0.5$, where the hopping parameters are defined in the same way as
in \cite{raghu}. Here, $E_F$=$-0.3 |t_1|$ for $25 \%$ hole doping.}
\vspace{-0mm} 
\end{figure}

\begin{figure}
\begin{center}
\vspace*{-2mm}
\hspace*{-0mm}
\psfig{figure=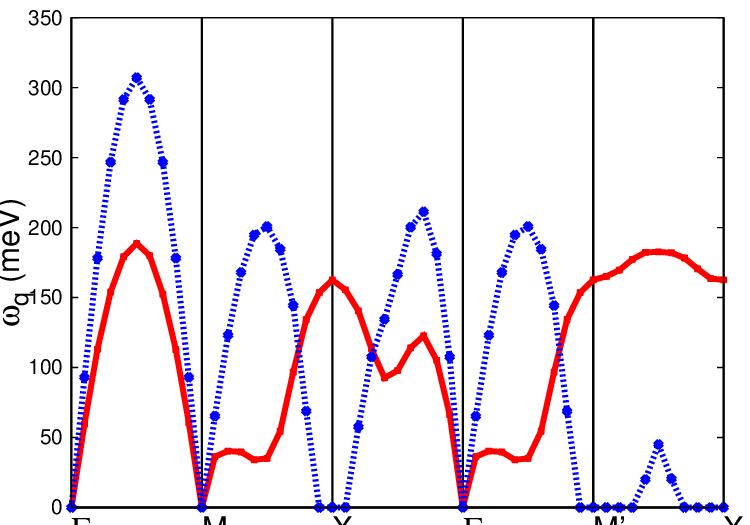,width=50mm,angle=0}
\end{center}
\caption{\label{fig9}
Spin wave dispersion for the same parameters as in Fig. \ref{fig8} (solid line)
and for nesting condition as in \cite{raghu} (dotted line). Here,
$\Delta$=3.0$|t_1|$, $J$=2.0$|t_1|$ and $|t_1|$=$200$
meV. The finite spin wave energy at M' indicates emergent F spin coupling.}
\vspace{-0mm} 
\end{figure}

\section{Modified version of two-band model of Raghu \textit{et al.}}
With the insight thus obtained, we consider a modified version of the two-band
model of Raghu \textit{et al.} which retains most
of their essential features, and simultaneously yields not only the two
circular hole pockets, but also appropriate orbital order, and also
evidence of emergent F spin couplings. The hopping parameters are defined in
the same way as in \cite{raghu} and their values are taken as $t_1$=$-1.0$,
$t_2$=$0.5$, $t_3$=$t_4$=$-0.5$. The corresponding Fermi surface and the spin
wave dispersion for $25 \%$ hole doping  are shown in Fig. \ref{fig8} and  Fig.
\ref{fig9}, respectively. The finite spin wave energy at the F zone boundary M'
implies the emergence of F spin coupling. Although nesting is no longer obtained
due to absence of electron pockets, we have  $n_{\alpha}\simeq n_{\beta}\simeq
0.75$ in
paramagnetic state, $n_{\alpha}\simeq 0.82$,$n_{\beta}\simeq 0.68$ in ($\pi,0$)
state and hence orbital order of correct sign is obtained in this model. This
suggests that an
additional $d_{xy}$ band may need to be included to simultaneously
understand all the important electronic Fermi surface features as observed in
ARPES experiments (including the elliptical electron pockets) and the spin wave
features as observed in INS studies.

\section{Conclusion}
In conclusion, we have shown that within a minimal two-orbital
itinerant-electron model with hopping anisotropy, the magnetically anisotropic
($\pi,0$) ordered SDW state is stabilized through the
generation of orbital order which optimizes both AF and F spin couplings.  We
have also shown that with the inclusion of inter-orbital Hund's coupling, the
spin stiffness further increases substantially due to the coupling of the two
magnetic sub-systems. The calculated spin-wave energy scale, nature of spin-wave
dispersion and presence of ferro-orbital order are in agreement with
experimental observations. Thus, the proclivity of iron arsenides towards
($\pi,0$) magnetic ordering may actively involve the orbital degree of freedom. 
We have also considered a two-band model taking hopping parameters similar to
Raghu \textit{et al.}\cite{raghu}, but with hopping anisotropy, which yields
appropriate orbital order, circular hole pockets, as well as emergent F
spin couplings. The presence of orbitally ordered state in our model
allows the possibility of exploring orbital fluctuations which have been seen in
recent experiments \cite{Kim}. Orbital fluctuations may play an
important role on the electron-paring mechanism \cite{Shimojima2} and can
induce $s++$ wave superconducting state \cite{Saito}, and therefore a
combination
of orbital and spin fluctuations as a possible mechanism for superconductivity
also needs to be explored in these materials.

Sayandip Ghosh acknowledges financial support from Council of Scientific and
Industrial Research, India.

\end{document}